# Challenges of Internet of Things and Big Data Integration

Zainab Alansari[1, 2], Nor Badrul Anuar[1], Amirrudin Kamsin[1], Safeeullah Soomro[2], Mohammad Riyaz Belgaum[2], Mahdi H. Miraz[2, 3] and Jawdat Alshaer[4]

[1] University of Malaya, Kuala Lumpur, Malaysia
`z.alansari@siswa.um.edu.my`

[2] AMA International University, Salmabad, Kingdom of Bahrain
`zeinab@ amaiu.edu.bh`

[3] Wrexham Glyndŵr University, Wrexham, UK

[4] Al-balqa Applied University, Jordan

**Abstract.** The Internet of Things anticipates the conjunction of physical gadgets to the Internet and their access to wireless sensor data which makes it expedient to restrain the physical world. Big Data convergence has put multifarious new opportunities ahead of business ventures to get into a new market or enhance their operations in the current market. considering the existing techniques and technologies, it is probably safe to say that the best solution is to use big data tools to provide an analytical solution to the Internet of Things. Based on the current technology deployment and adoption trends, it is envisioned that the Internet of Things is the technology of the future; while to-day's real-world devices can provide real and valuable analytics, and people in the real world use many IoT devices. Despite all the advertisements that companies offer in connection with the Internet of Things, you as a liable consumer, have the right to be suspicious about IoT advertisements. The primary question is: What is the promise of the Internet of things concerning reality and what are the prospects for the future.

## 1   Introduction

Convergence between wireless communications, Digital electronic devices, and Micro-electro-mechanical systems (MEMS) technologies led to the rise of the Internet of Things. Internet-connected objects like computers, smartphones, tablets and Wi-Fi devices, sensors, wearable devices and household appliances are all the objects of the IoT components [1] and considered as "Things". The Internet of Things means the production of tremendous amount of data and a collection of substantial different data bulk that has not seen so far. Big data management and generating smart data are the research interests of the companies which produces these data. Without the application of big data analytics, the vast volume of data generated by the Internet of Things adds to the overhead of any organization and considered to be one of the most significant obstacles towards the deployment of this technology [2]. In other hand, organizations must know what to do with the massive amount of data that collected. The explosive growth of Internet use, along with smartphone and social programs and machine-to-machine (M2M) communication, has revolutionized big data [3].

The primary challenge is the design of a model to analyze big data. In other words, we need to change our point of view about the blocks created by the Internet of Things [4]. Instead of looking at data as a data warehouse, we should look at the supply chain. Since the tools are enabling the extraction of numerous unstructured and new data sources, the lack of sufficient data issue will reveal soon. Therefore, must overcome the following two fundamental problems [5]. Firstly, not to miss the data that truly needed and secondly, make sure not to spend time on unnecessary data. Despite the supply chain, organizations can fill in existing gaps in any way they need [6]. To this end, companies can take advantage of these three strategies:

- Design interfaces for applications that created before.
- Request help from partners or third parties who can provide the required data.
- Generate data with the commission of the physical environment around the business.

Apparently, getting the accurate data is not the only problem with organizations. The other challenge is to acquire the necessary skills in the field of analytical analysis to deal with the big data. Traditional skills in the domain of data analysis on the Internet of Things have not been practical. Companies need people who know analytics, as well as understanding the new meaning of the initial data for a specific industry. Along with the growing trend in data generation and analysis required for these big data, organizations are forced to prepare devices that connect customers and objects at any time and any place [7].

One of the critical infrastructures required by active companies in the Internet of Things is having the culture of data-driven decision-making. The Internet of Things, in essence, provides a flow of accurate data derived from the real world. These data must go beyond the process of transforming data into information, then knowledge and awareness, and ultimately wisdom, using traditional analytical skills in this area to be meaningful. For example, in the field of agriculture, an expert scientist needs to know how much irrigation required to produce a product under different climatic conditions [8].

In intermittent IoT, the ability to collect weather information, field, and product information is automatic and accurate. However, when the data collected, the actions to be taken on data depends on the expert opinion. Coupled with the growing amount of data and analysis needed, companies need to be ready for a range of devices that connect consumers and objects at any point and any time. Those who accept the philosophy of data supply chain will go through these waves of information without dwelling in detail. Apart from the cases mentioned, the investment required in the field of sensors, analytical capabilities, and data security and support are among the other obstacles faced by the technology of the Internet of Things [9].

Hitherto, considering the existing techniques and technologies, it is probably safe to say that the best solution is to use big data tools to provide an analytical solution to the Internet of Things. Based on the current technology deployment and adoption trends, it is envisioned that the Internet of Things is the technology of the future; while today's real-world devices can provide real and valuable analytics, and people in the real world use many IoT devices. Despite all the advertisements that companies offer in connection with the Internet of Things, you as a liable consumer, have the right to be suspicious about IoT advertisements. The primary question is: What is the promise of the Internet of things concerning reality and what are the prospects for the future?

The core value of an IoT system is the ability to analyze the data needed and achieve practical and useful insight without making any mistake. Hence for two reasons, creating a communication medium is not easy. Firstly, it must have the ability of scalable analysis. Secondly, it must make comprehensive usage of this ability regarding the volume and speed of the IoT devices that generate their data [10]. In this research, we discussed different solutions to help everyone to stay away from a series of issues on how to develop an ideal analysis of big data.

## 2   Internet of Things and Its Impact on Big Data

Nowadays, with the help of the Internet of Things technology, the ability to connect each object to the network is provided. The Internet of Things offers a chain of connected people, objects, applications, and data over the Internet for remote control, interactive, services integration and management. Hence this network is overgrowing; we need a platform that can collect and store the data generated by IoT devices. Some of the advanced Internet of Things services require a mechanism to collect, analyze and process raw data from sensors to be used as operational control information. Some types of sensor data may have very high volumes because of the significant number of sensors in the Internet of Things ecosystems. Possibly, we would see a data flood coming from these devices. Accordingly, we need new technologies or architectural patterns in the area of data collection, storage, processing, and data retrieval [11].

Databases designed and implemented to work with the Internet of Things have their specific conditions and characteristics. The proliferation of NoSQL technologies can be considered as an indication that the management of the Internet of Things requires the use of novel approaches in administering and utilizing databases. The provision of cloud computing platforms based on the internet of things eases the opportunities to enter this arena and take advantage of its achievements and services for many businesses of various dimensions [12]. Moreover, there will be a need for IoT data analysts of an acquaintance and entry into the fascinating world of big data. The Internet of things affects people, processes, data, and things.

•      People: More objects can be monitored and controlled, and subsequently increased individual's abilities.

•      Processes: Users and more machines will be able to interact with each other in real time. Therefore, very complex tasks can finish in less time as the percentage of engagement and participation in doing a job is far more significant.

- Data: The ability to collect data at a higher frequency and reliability provided which can lead to a correct decision making.
- Things: the ability to control things more accurately. Therefore, the value of objects such as mobile devices will be more and can help with much more than the current situation.

Big data convergence, super-efficient networks, social media, low-cost sensors, and a new generation of advanced analytics have provided countless new opportunities for business enterprises to use them to either enter a new market or strengthen their operations in the current market. The Internet of Things is one of these new markets that can provide countless opportunities for businesses in different fields. Significant changes happen by a slight difference, and the Internet of Things can be the source of millions of changes in different areas over the next few years. Consider the Internet of Things as one of the causes for generating data which its impact on IT infrastructure and the use of advanced methods in data analysis are among its exceptional and vital opportunities in this regard [13].

A collection, preparation, and analysis of large volume of data will not be an easy task. Firstly, the amount of data can be doubled in several months and secondly, the gendered nature of this kind of data has its particular complexities. The variety in the template or the format of this type of data is extensive and often includes hundreds of pseudo-structured forms or unstructured formations. Most importantly, to achieve a broad view of the sensor data, it should be possible to analyze and manage every structured and unstructured data. An analysis based on a specific data format can significantly limit the created potential insight. Data analysis, regardless of its composition, is centralized and side by side, which provides a comprehensive analytical perspective to decision makers of business enterprises.

It indicates the consideration of the limitations and stock of traditional enterprise data and current business intelligence software and design. Organizational data warehouses are not able to focus on unstructured data. Accordingly, we need to look for solutions that enable unstructured data storage and analysis [14]. If we want to convert the unstructured data into a specific structure by defining a particular structure and using relational database tables, we will lose time. Consequently, that will surely be possible with the condition of not having the technical limitations. The use of any technology to create an analytical infrastructure that has some limitations can reduce the ability to analyze and, in practice, minimizes the potential for possible value creation.

Analysis of big data with the help of related technologies can be one of the leading actors in this field. Analyzing big data in some cases can help us:

- Combine, integrate and analyze all structured, semi-structured and unstructured data regardless of source, type, size, and format.
- A quick and cost-effective analysis of the high volume data to create an appropriate insight into a decision making process.

The Internet of Things has promised to influence various industries from insurance companies and banks to telecom and other business enterprises. Organizations need to modify data analysis methods so that they can collect, clean, prepare and analyze RFID sensor and tag data in the shortest time possible. Deciding on the actual data and in the shortest reasonable time is the confidentiality of a business firm in today's highly competitive environment. With a proper big data management and the creation of an appropriate atmosphere for their analysis, an organization's vision for proper decisions making increases [15].

## 3    Benefits of IoT based on Big Data

In literature, various structures for big data analysis and IoT proposed, which can manage the challenges of storage and analysis of high volume data from intelligent buildings. The first presented structure consists of three main components which are big data management, IoT sensor, and data analysis. These analyzes use are in the real-time management of oxygen level, dangerous gases/soot and the amount of ambient light in smart buildings. In addition to smart building management, IoT devices and sensors for receiving traffic information can be used in real time traffic management with low cost and examine the strengths and weaknesses of existing traffic systems.

In smart city management, the big data used in the analysis of data which obtained from different sensors such as water sensors, transportation network sensors, monitoring devices, smart home sensors and smart car park sensors.  These data are generated and processed in a multi-stage model and ultimately reached a decision-making stage. These steps are data production, data collection, data integration, data categorization, data processing and decision making [16].

Sometimes it is essential to pay attention to the concepts of web technology in particular proposed framework to investigate the analytical results obtained from the big data in the Internet of Things. In the literature, this topic has devised, and a conceptual framework has been proposed consisting of 5 layers:

•    Data Collection layer - collected data from various sources, the input layer is the proposed framework.

•    Extract-transform-load (ETL) layer - provides the ability to change the format of information received from different types of sensors into a defined format.

•    The semantic reasoning rules layer - an inference engine that acts on the information received from the ETL layer

•    Learning layer - From the data tailored to the existing extraction data, extract the various specifications and attributes, and finally, Machine learning-based models provided.

•    Action layer - executes a set of predefined actions by the outputs of the learning layer.

Other applications of IoT help with geographic information analysis, cloud computing flow processing, big data analysis, and storage, cloud computing security, clustering mechanisms, health, privacy security, performance evaluation of monitoring algorithms, manufacturing systems, and energy development [17].

## 4 Big Data Collection and Storage

Numerous protocols make it possible to receive events from IoT devices, especially at lower levels. It does not matter if the device connected to a Bluetooth network, cellular network or Wi-Fi, or it communicates through a hardware connection, it is enough to send a message from a broker using a defined protocol. One of the most popular protocols for large IoT applications is MQ Telemetry Transport (MQTT). MQTT refers to the transmission of messages through remote sensing and queuing which is an M2M IoT connection. This protocol designed as a very lightweight request/ response (point-to-point) messaging transfer. MQTT is practical and useful for connecting to distant locations that require low memory or low network bandwidth; For example, this protocol used in sensor communication through a satellite link with a broker in dial-up connections with healthcare providers at different times and in a range of home automation and small devices. Its design principles are to minimize network bandwidth and resource requirements, and at the same time, it also guarantees trust and confidence in the message delivery [18].

There are also other alternatives, such as the limited application protocol, XMPP, and other protocols. Constrained Application Protocol (COAP) is a software protocol that used in straightforward electronic devices which provides communication through the Internet interactively. Constrained RESTful Environments (CoRE) group along with Internet Engineering Force (IETF) did the main standardization work of this protocol.

Extensible Messaging and Presence Protocol (XMPP) is a communication protocol for the Extensible Markup Language (XML). The XMPP enables exchanging close to real time between structured data but expandable between either any two or more network entities.

We recommend starting with MQTT due to its the availability and extensive coverage and the availability of a large number of client applications and open source brokers unless having a convincing reason to choose another protocol. Mosquitto is one of the MQTTs most widely used open source and will be the definitive choice for applications; if this concept should be proved based on limited budgets and want to avoid the cost of dedicated devices, the fact that Mosquitto is open-source brokerage is precious. Regardless of what protocol to choose, it will eventually receive messages that represent events and observations from devices connected to the Internet.

As long as the message received by a broker (Mosquitto), you can send it to the analytics system. The best way is to store source data before any transfer or making any changes to them; this is, of course, worthwhile when debugging problems occurs at the conversion stage. There are several ways to store IoT data which many uses Hadoop and Hive in their projects. Recently researchers are and successfully working with NoSQL databases such as Couchbase. Couchbase provides a right combination of high performance and low latency indicators. Couchbase is a Document-Oriented Database which lacks a layout that covers a significant amount of data and also can add a variety of new events [19] quickly.

Direct data writing on HDFS is another good solution, especially if using Hadoop and batch analysis is considered as part of the analytics workflow. For writing source data in a permanent storage location, the direct code can be added manually to the message broker at the level of the IoT protocol (for example, if you use MQTT, Mosquitto Broker). The other way is to attach messages to a medium-sized middleware broker like Apache Kafka and use different Kafka users to transfer messages to different parts of the system [20].

One of the established patterns is to place messages in Kafka in two user groups based on the subject, where one of the groups writes raw data to their permanent storage, while another, transmits data into a real-time processor engine like Apache Storm. If Storm is used instead of Kafka, a Bolt processor can install in topology, which does nothing except sending messages to a permanent storage location. If MQTT and Mosquitto are used, sending direct messages to the Apache Storm topology via the MQTT's spout source is an easy way to link things together.

## 5 Conclusion

The development of IoT devices, smartphones, and social media provides decision makers with opportunities to extract valuable data about users, anticipate future trends and fraud detection. With the creation of transparent and usable data, big data can create the organizations' values, make the changes clear and expand their performance. The use of data generated from the IoT and the analytical tools creates many opportunities for organizations. These tools use predictive modeling technologies, clustering, classification to provide data mining solutions.

IoT improves the decision-making habits of decision-makers. The emergence of IoT and related technologies, such as cloud computing, provides the ability to remove data sources in different domains. Typically, any data is considered useful in the domain itself, and data on shared domains can be used to provide different strategies. Machine learning, deep learning, and artificial intelligence are key technologies that are used to provide value-added applications along with IoT and big data in addition to being used in a stand-alone mode. Before the advent of IoT and cloud computing, the use of these technologies was not possible due to the high amount of data and required computational power. Different data analysis platforms, Business intelligence platforms and analytical applications are emerging platforms that have been introduced to help industries and organizations in transforming processes, improving productivity, and the ability to detect and increase agility.

It is anticipated that the speed of technological progress in the next ten years, will be equal to the past thirty years. Therefore, we have to use all our efforts to update our lives to the Internet of Things technology regarding hardware and software.

**Acknowledgment**

This work is supported by CTRG Research Group of the College of Computer Studies, AMA International University, Kingdom of Bahrain.